\documentclass[acmsmall]{acmart}
\AtBeginDocument{%
  }

\setcopyright{acmlicensed}
\copyrightyear{2025}
\acmYear{2025}
\acmDOI{XXXXXXX.XXXXXXX}

\acmJournal{JACM}
\acmVolume{37}
\acmNumber{4}
\acmArticle{111}
\acmMonth{8}

\begin{document}

\title{Bounded Minds, Generative Machines}
\subtitle{Envisioning Conversational AI that Works with Human Heuristics and Reduces Bias Risk}
\author{Jiqun Liu}
\email{jiqunliu@ou.edu}
\orcid{0000-0003-3643-2182}
\affiliation{%
  \institution{The University of Oklahoma}
  \city{Norman}
  \state{Oklahoma}
  \country{USA}
}




\renewcommand{\shortauthors}{Trovato et al.}

\begin{abstract}
Conversational AI is rapidly becoming a primary interface for information seeking and decision making, yet most systems still assume idealized users. In practice, human reasoning is bounded by limited attention, uneven knowledge, and reliance on heuristics that are adaptive but bias-prone. This article outlines a research pathway grounded in bounded rationality, and argues that conversational AI should be designed to work with human heuristics rather than against them. It identifies key directions for detecting cognitive vulnerability, supporting judgment under uncertainty, and evaluating conversational systems beyond factual accuracy, toward decision quality and cognitive robustness.

\end{abstract}



\received{20 February 2007}
\received[revised]{12 March 2009}
\received[accepted]{5 June 2009}

\maketitle

\section*{Bounded Rationality in Conversational AI}
Conversational AI systems are entering the mainstream as a primary interface for information seeking and decision making. Yet most systems, evaluations, and implicit user models still presume idealized users who can carefully weigh evidence and notice implausible claims. In reality, human cognition is \textit{bounded} by limited time and attention, uneven domain knowledge, and reliance on heuristics that evolved to support fast judgments under uncertainty~\cite{tversky1992advances}. Heuristics such as anchoring, availability, confirmation, authority, and scarcity are not failures of reasoning but adaptive mental shortcuts. When paired with fluent, personalized, and confident generative responses, however, these shortcuts become a double-edged mechanism that can systematically skew judgment.

Researchers in human-computer interaction, cognitive science, and behavioral economics have long studied these biases, and there is growing recognition in AI ethics and safety communities as well~\cite{draws2021checklist}. However, mainstream design practices and system evaluations still emphasize correctness, relevance, and narrowly-defined utility while leaving cognitive impact underexplored~\cite{liu2023behavioral}. In multi-turn dialogue, a single confident response can anchor users, vivid anecdotes can dominate base rates, repeated agreement can reinforce expectations and misbeliefs, and even neutral layout or ordering choices can introduce framing effects that alter trust and preference.

The risk is not merely that conversational AI fails to mitigate these biases; it is that systems can exploit them for scalable behavioral manipulation, which aligns with the unacceptable-risk framing under the EU AI Act~\footnote{See https://www.europarl.europa.eu}. In advertising, persuasion, and political microtargeting, systems already optimize engagement by aligning with predictable vulnerabilities~\cite{carroll2023characterizing}. As generative models become more adaptive and personalized, subtle steering becomes easier to produce and harder to detect.

The challenge, then, is not simply higher factual accuracy. It is to build conversational systems that recognize how users actually reason through heuristics that can be adaptive or distorting depending on context, individual differences, and system choices. An emerging line of work, sometimes described as \textit{Machine Psychology}, studies how generative models exhibit and interact with cognitive shortcuts and how systems might support better reasoning rather than amplify bias~\cite{rahwan2019machine}. This motivates a research pathway that connects design, evaluation, and governance to reduce AI harms and bias risks in practice~\cite{magooda2023framework}.

\paragraph{Why this becomes a research pathway now?}
Bounded rationality has long been studied, but its implications become substantially more urgent with large language models (LLMs) and agentic conversational systems being widely deployed. Unlike earlier information systems, conversational AI can adapt natural language turn by turn and influence reasoning over longer horizons, amplifying cumulative effects of framing, anchoring, and heuristic reinforcement. Over the next decade, this makes cognitive impact a first-order design concern rather than a peripheral issue.

\begin{figure}
    \centering
    \includegraphics[width=0.8\linewidth]{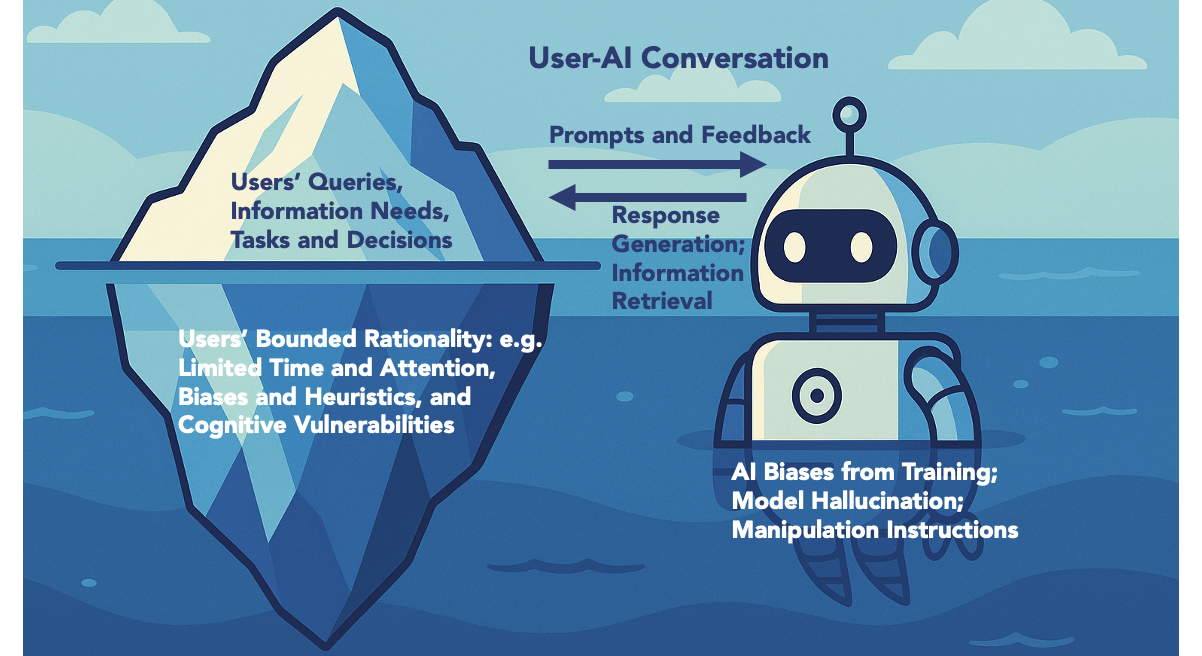}
    \caption{Challenge in User-AI conversation: System responds to the explicitly issued prompts, but cannot detect underlying bounded rationality or support the implicit human heuristics "under the water".}
    \label{fig:flow}
\end{figure}

\section*{Designing for Heuristics, Not Against Them}
This work does not argue for a single design solution. Instead, it highlights recurring cognitive dynamics that define a broader research pathway for conversational AI as heuristics unfold across multi-turn interaction.
Once we recognize that people rely on heuristics rather than mathematically rational utility calculation, the research challenge is to understand how these shortcuts behave when responses are generated fluently, adaptively, and at scale across varying tasks. What once appeared as subtle laboratory effects can accumulate across everyday exchanges, where small shifts in wording, sequencing, and emphasis reshape judgments, comparisons, and preferences.

According to Prospect Theory~\cite{tversky1992advances}, people evaluate outcomes relative to \textit{reference points} and weigh losses more than equal gains. In conversational settings, reference points are often established implicitly through language, ordering, and tone, and early framings can persist across turns. Over time, path-dependent agreement can narrow exploration even without introducing factual errors.


The structure of option sets also reshapes outcomes. When comparisons are generated on the fly, dominated alternatives may appear and quietly steer choices. For example, a student asking an LLM for study resources might see a free open textbook, a reasonably priced course, and an overpriced ``platinum package'' with little added value; the decoy makes the paid course look like the safe and balanced choice. Similar dynamics arise with anchoring, vivid anecdotes, authoritative tone, scarcity cues, and confirmation-seeking language, which can co-occur and reinforce one another in conversational information retrieval and recommendation~\cite{liu2023behavioral}.

These dynamics explain why surface-level accuracy cannot be the only standards for judging conversational AI. A response may be correct yet still entrench loss aversion, amplify overconfidence, or channel decisions through hidden framing. The research question shifts from whether an answer is correct to whether an interaction supports sound reasoning within bounded cognitive limits: does it broaden perspectives, calibrate uncertainty, and reduce vulnerability rather than compound it? Because interactions unfold turn by turn and are context-dependent, this pathway calls for design and evaluation centered on cognitive impact in addition to quality-focused effectiveness.

\section*{Detecting and Intervening in Bias Risks}
Conversations are where reasoning develops between human and machine, and also where it can quietly bend. This pathway requires monitoring how exchanges unfold and identifying signals of bias risk: narrow framing, unbalanced emphasis on benefits versus costs, overconfident claims without uncertainty, comparisons that hide baselines or include redundant options, and rapid user acceptance without examining evidence. Such signals suggest measurable features that future systems could track, including viewpoint diversity via source entropy~\cite{lewis2020retrieval}, calibration failures in expressed probabilities~\cite{guo2017calibration}, and turn-level traces that enable later audit~\cite{magooda2023framework}. Additional behavioral and physiological signals may help in some settings, but generalizable, privacy-aware signals remain a core research challenge.

Once signals are detected, systems could explore lightweight interventions before presenting a final response: surfacing a clear baseline, framing outcomes symmetrically in gain and loss terms, and representing uncertainty as ranges rather than single numbers~\cite{guo2017calibration}. Retrieval can supply base rates and counterexamples~\cite{lewis2020retrieval}. A dominance check can flag or remove decoys, and an anchoring guard can introduce a second estimate after baselines are shown so the first number does not dominate judgment~\cite{yao2023react}. These interventions can be implemented through structured outputs, function calls, and lightweight critics that review drafts before answers are returned~\cite{ouyang2022training}.

Illustrative scenarios can clarify the goal. In health advice, a vivid patient story presented first can dominate evidence from clinical trials; pairing anecdotes with prevalence rates can rebalance judgment. In e-commerce, a high-priced decoy can make a mid-range product appear reasonable; detecting and labeling dominated options reduces hidden steering. Similar failures can arise in legal guidance and news summarization through redundant options, problematic framing, or anchoring order. These cases show how heuristics can turn from shortcuts into vulnerabilities and why small interventions may keep conversations more balanced under bounded rationality.

Evaluation also has to evolve. Accuracy and topical relevance remain necessary but are no longer sufficient. Systems should be tested on whether interactions improve calibration after uncertainty is shown~\cite{guo2017calibration}, broaden perspectives, and reduce susceptibility to dominated options and decoys~\cite{magooda2023framework}. Scenario sets that vary framing, priors, and time pressure can stress-test interventions, while online experiments can compare policies through stability of choices and reduced repeat queries~\cite{ouyang2022training}. Counterfactual logging and interleaving can evaluate alternative baselines safely, and tuning can incorporate penalties when high-risk turns pass without intervention~\cite{magooda2023framework,ouyang2022training}.

\paragraph{Open challenges.}
Detecting cognitive vulnerability in real time is subtle, context-dependent, and heterogeneous across users and domains. Interventions must balance effectiveness with intrusiveness, since excessive safeguards may increase cognitive loads and reduce trust or usability. Integrating detection, intervention, evaluation, and governance into a coherent conversational pipeline raises unresolved questions about transparency and generalizability. These challenges are central to the research pathway rather than solved engineering tasks.

\section*{From Principles to Practice}
Turning these ideas into practice does not require reinventing conversational AI, but it does require reframing priorities. Many building blocks exist, yet their effective integration and evaluation remain open research problems. Function calls, structured outputs, and long-context modeling can support baseline surfacing and uncertainty representation; retrieval-augmented generation and reasoning can supply diverse perspectives with grounded sources~\cite{lewis2020retrieval}; and lightweight critics can flag missing baselines or overconfident tone~\cite{ouyang2022training}. Recent model developments make these capabilities increasingly feasible, but feasibility alone does not guarantee cognitive benefit and safety.\footnote{See OpenAI, ``Introducing o3 and o4-mini,'' \url{https://openai.com/index/introducing-o3-and-o4-mini/}}

\paragraph{Open research questions and roadmap.}
This pathway suggests several actionable questions: (1) How can systems detect moments of cognitive vulnerability reliably across tasks and populations? (2) Which interventions, such as contrastive framing, baseline surfacing, or uncertainty representation, work best under different heuristics and contexts? (3) How should success be evaluated beyond accuracy, especially in calibration, stability of judgment, and resistance to manipulation? (4) What governance mechanisms and audit trails are needed to make such systems accountable at scale?

Evaluation can be reframed around these goals. Instead of asking only whether an answer is correct, we can ask whether interaction improves calibration, reduces anchoring, and broadens perspectives~\cite{guo2017calibration,magooda2023framework}. Scenario sets that vary framing, priors, and time pressure can test robustness, while online experiments can compare policies through stability of choices and reduced susceptibility to decoys~\cite{ouyang2022training}. Over time, such measures can complement accuracy and latency with benchmarks for decision quality and cognitive robustness.

Practical safeguards also depend on governance and transparency. \textit{Cognitive-impact reviews} can become part of release practice alongside privacy and security checks. Policy and model cards can disclose guardrail scope and known limitations, while audit trails can log which options were shown, what baselines were included, and which interventions were triggered~\cite{magooda2023framework}. Practical audit trails can include retrieval queries, ranking choices, cited spans, and the system's internal plan behind high-stakes answers.\footnote{See Anthropic Docs, ``Computer Use Tool,'' \url{https://docs.anthropic.com/en/docs/agents-and-tools/tool-use/computer-use-tool}} Such documentation may feel burdensome, but it enables reproducible traceability for accountability.

Taken together, these directions outline a credible and timely research pathway for building conversational AI systems and evaluation pipelines that work with bounded human minds, support robust reasoning, and reduce the risk of hidden harm and cognitive behavioral manipulation.

\section*{Author Bio}
Jiqun Liu is an Associate Professor at the University of Oklahoma, Norman, OK, USA.

\bibliographystyle{ACM-Reference-Format}
\bibliography{sample-base}
\end{document}